\documentclass[final,5p,times,twocolumn,sort&compress]{elsarticle}

\usepackage[numbers]{natbib}

\begin{document}

\begin{frontmatter} 

\title{Suppressing dynamical diffraction artefacts in  differential phase contrast scanning transmission electron microscopy of long-range electromagnetic fields via precession}

\author[Monash]{T. Mawson}
\author[JEOL]{A. Nakamura}
\author[MCEM,Monash]{T.C. Petersen}
\author[Tokyo,JFCC]{N. Shibata}
\author[Furukawa]{H. Sasaki}
\author[Monash]{D.M. Paganin}
\author[Monash]{M.J. Morgan}
\author[Monash]{S.D. Findlay \corref{cor1}}
\ead{scott.findlay@monash.edu}

\cortext[cor1]{Corresponding author}

\address[Monash]{School of Physics and Astronomy, Monash University, Victoria 3800, Australia}
\address[JEOL]{JEOL Ltd., Akishima, Tokyo 196-8558, Japan}
\address[MCEM]{Monash Centre for Electron Microscopy, Monash University, Victoria 3800, Australia}
\address[Tokyo]{Institute of Engineering Innovation, School of Engineering, University of Tokyo, Tokyo 113-8656, Japan}
\address[JFCC]{Nanostructures Research Laboratory, Japan Fine Ceramics Center, Nagoya 456-8587, Japan}
\address[Furukawa]{Furukawa Electric Ltd., Yokohama 220-0073, Japan}

\begin{abstract}
In differential phase contrast scanning transmission electron microscopy (DPC-STEM), variability in dynamical diffraction resulting from changes in sample thickness and local crystal orientation (due to sample bending) can produce contrast comparable to that arising from the long-range electromagnetic fields probed by this technique. Through simulation we explore the scale of these dynamical diffraction artefacts and introduce a metric for the magnitude of their confounding contribution to the contrast. We show that precession over an angular range of a few milliradian can suppress this confounding contrast by one-to-two orders of magnitude. Our exploration centres around a case study of GaAs near the [011] zone-axis orientation using a probe-forming aperture semiangle on the order of 0.1 mrad at 300 keV, but the trends found and methodology used are expected to apply more generally.
\end{abstract}


\end{frontmatter}

\section{Introduction}

Transmission electron microscopy (TEM) is well suited to the problem of mapping and measuring electromagnetic fields inside materials, being capable of high spatial resolution and being sensitive to these fields since they scatter the electron beam. In classical terms, this scattering results from the Lorentz force of the electromagnetic field acting on the probe electrons. In wave-optical terms, it results from the phase profile imparted upon the electron wavefield by the sample's electromagnetic potential via the Aharonov-Bohm effect \cite{aharonov1959significance}. All elastic-scattering techniques in electron microscopy---including holography, Fresnel and Foucault imaging, and ptychography---derive from this.

One conceptually-simple technique for electromagnetic field mapping is differential phase contrast (DPC), which consists of mapping the deflection of an electron probe as it is scanned across the sample. This has long been used to map out magnetic domain structure in materials \cite{CBWF1,CPD1,UZ1,URDHHWZ1,matsumoto2016direct}, and more recently to probe long-range electric structures in materials, including polarization distributions \cite{LSJWWSZ1,SFKSKI1,BHLRBZ1} and the built-in field at p-n junctions \cite{SFSMSKOMI1,haas2019direct}. While the rigid deflection model may be an over-simplification for quantitative work \cite{clark2018probing}, for phase objects the first moment or centre of ``mass'' (read intensity) of the diffraction pattern is known to be equal to the gradient of the phase profile imparted by the sample convolved with the probe intensity distribution \cite{WC1,Muller_2014_KBSGLVZSR,Lubk_2015_Z}. The first moment can be measured to high precision with fast-readout pixellated detectors \cite{Muller_2014_KBSGLVZSR,tate2016high} or to a good approximation using segmented detectors \cite{Close_2015_CSF,lazic2016phase,taplin2016low}.

\begin{figure*}[htb!]
	\centering
		\includegraphics[scale=0.95]{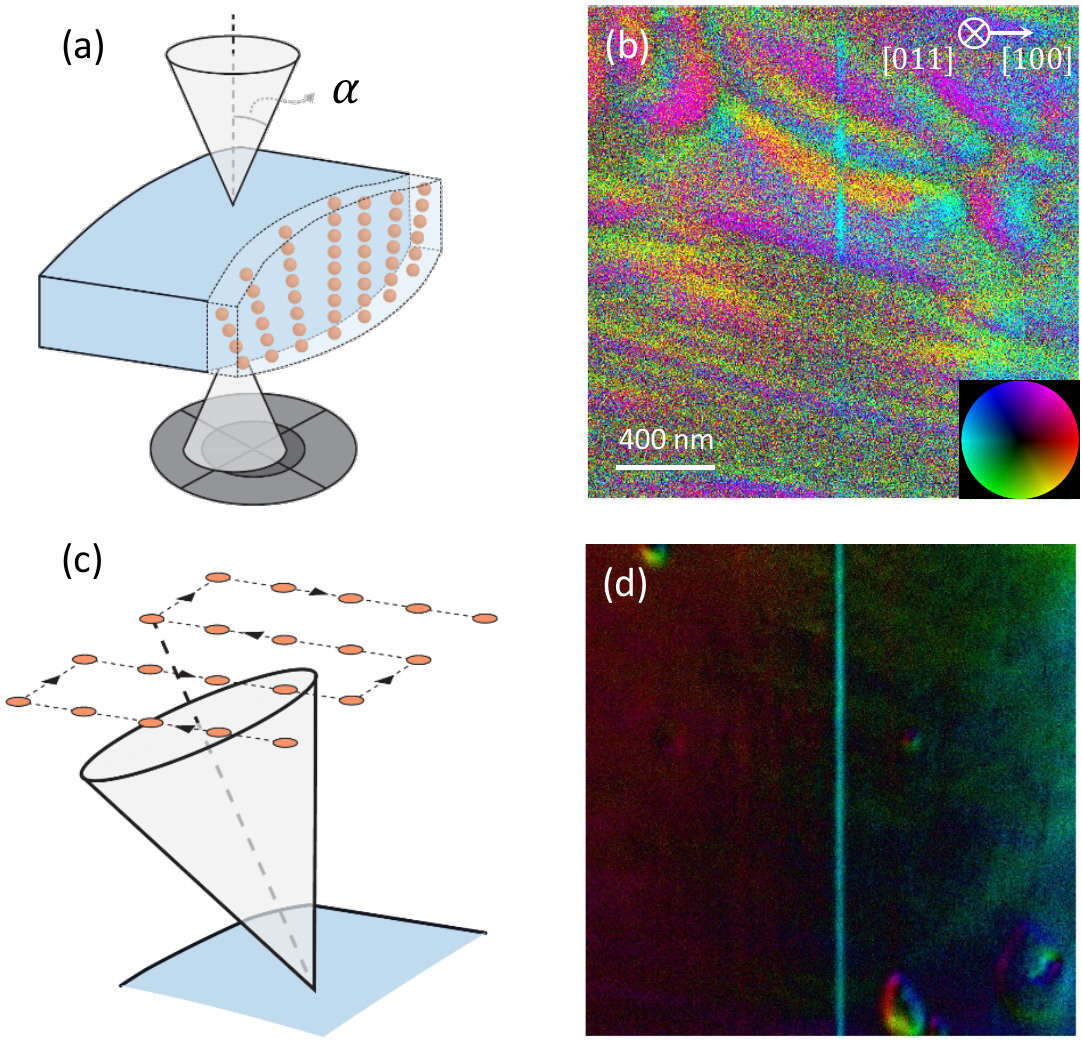}
	\caption{(a) Schematic of a convergent STEM probe, with probe-forming aperture semiangle $\alpha$, incident upon a crystal with greatly exaggerated thickness variation and sample bending (depicted through the change in orientation of the columns of atoms---orange spheres---in the partially transparent cutaway). The scattered beam is shown falling upon a segmented detector. (b) Experimental DPC-STEM image from a GaAs sample aligned near the [011] zone axis orientation and containing a p-n junction running vertically through the centre of the field of view. The image is shown as a vector colour map, with colour denoting direction and value (intensity) denoting magnitude as per the inset colour wheel. The intensity scale has been partially saturated to make the p-n junction more visible. Note that we depict the first moment as estimated on the segmented detector rather than the electric field that would produce such a deflection. At the p-n junction, such a built-in electric field is truly present. However, the more horizontally-oriented swathes of colour are not thought to reflect true long-range fields but rather variations in dynamical diffraction condition due to variations in thickness and sample bending across the field of view. (c) Schematic showing the tilt angle pattern for precessing the incident beam orientation relative to the sample. (d) Experimental DPC-STEM image resulting from precession averaging over an $11 \times 11$ square grid of tilt angles (the image in (b) being from one such orientation) in 0.05 degree increments. Results from this data set were previously presented in Ref. \cite{nakamura2017differential}.}
	\label{F1}
\end{figure*}

However, contributions to the DPC signal may arise from factors which do not reflect the long-range electromagnetic field distribution. Dynamical diffraction, i.e.~multiple electron scattering through depth, means that real samples are rarely the pure phase objects assumed in DPC analyses \cite{maclaren2015origin,krajnak2016pixelated,taplin2016low,cao2017theory}. Moreover, as the probe is scanned across the field of view variations in dynamical diffraction arising from variations in sample thickness and the local sample orientation (i.e.~of the atomic planes) due to sample bending---depicted in Fig.~\ref{F1}(a)---can lead to a variation in the first moment of the diffraction pattern that does not reflect the long-range electromagnetic fields. Though in principle  conveying genuine information about the sample structure, we will refer to such contributions as dynamical diffraction artefacts since they confound the reliable interpretation of DPC maps as depicting long-range electromagnetic fields.

As an example, Fig.~\ref{F1}(b) shows an experimental DPC-STEM image of a GaAs sample containing a p-n junction \cite{nakamura2017differential}. The junction is perceptible as the fuzzily-straight, vertical, cyan stripe across the centre of the image. However, other DPC contrast, primarily consisting of horizontal swathes of colour indicating variously-oriented deflection, is also evident throughout the field of view. Though often involving a larger beam deflection than that caused by the p-n junction (the intensity scale in the figure has been partially saturated to make the p-n junction more visible), this other contrast is not thought to result from true long-range electric fields but rather from variations in thickness and/or local sample orientation. We also note that Fig.~\ref{F1}(b) contains screw-type and edge-type topological defects in the direction of electron deflection; the former class of defect corresponds to points where all hues on the colour wheel converge, while the latter correspond to lines across which the colour-wheel hue rotates by 180 degrees \cite{SethnaBook}.

Strategies to reduce the contribution from dynamical diffraction artefacts (i.e. beam deflection arising from scattering from the atomic electric fields as opposed to the long-range electromagnetic fields of interest) include judicious choice of sample orientation \cite{SFSMSKOMI1,haas2019direct}, isolating the central beam \cite{cao2017theory}, and excluding the contribution from the middle portion of the central beam \cite{krajnak2016pixelated}. The strategy we will concentrate on here is precession \cite{nakamura2017differential}. Precession---often conceptually described as precessing the sample about the optical axis but more usually implemented by rocking the beam pre-specimen (see Fig.~\ref{F1}(c)) and de-rocking it post-specimen---has been shown to make the resultant diffraction patterns more kinematical, i.e.~to reduce dynamical diffraction effects (reviews can be found in Refs. \cite{eggeman2012precession,midgley2015precession}). Originally used for solving crystallographic structures, the technique has since been used to extend the reliability of orientation mapping of grains in polycrystals \cite{midgley2015precession}, to improve the precision of strain mapping \cite{rouviere2013improved}, and, as recently proposed by Nakamura \emph{et al.} \cite{nakamura2017differential}, to improve the interpretability of DPC mapping of electromagnetic fields. Nakamura \emph{et al.}'s proof-of-principle result is reproduced in Fig.~\ref{F1}(d) using the same intensity scale as the image in Fig.~\ref{F1}(b). The confounding contrast has been largely suppressed, significantly improving the visibility of the p-n junction.

In this paper we explore how the dynamical diffraction artefacts arising from variations in the orientation of lattice planes and in sample thickness impact on the DPC signal through simulations for the same sample and broadly similar experimental conditions to those of Fig.~\ref{F1}. Developing metrics for this impact, we quantify the effectiveness of precession for suppressing these artefacts. We also explore the relative impact of orientation variation versus thickness variation. We round out our analysis by comparing against experimental data.

\section{Analysis method}

\begin{figure}[htb!]
	\centering
		\includegraphics[scale=0.9]{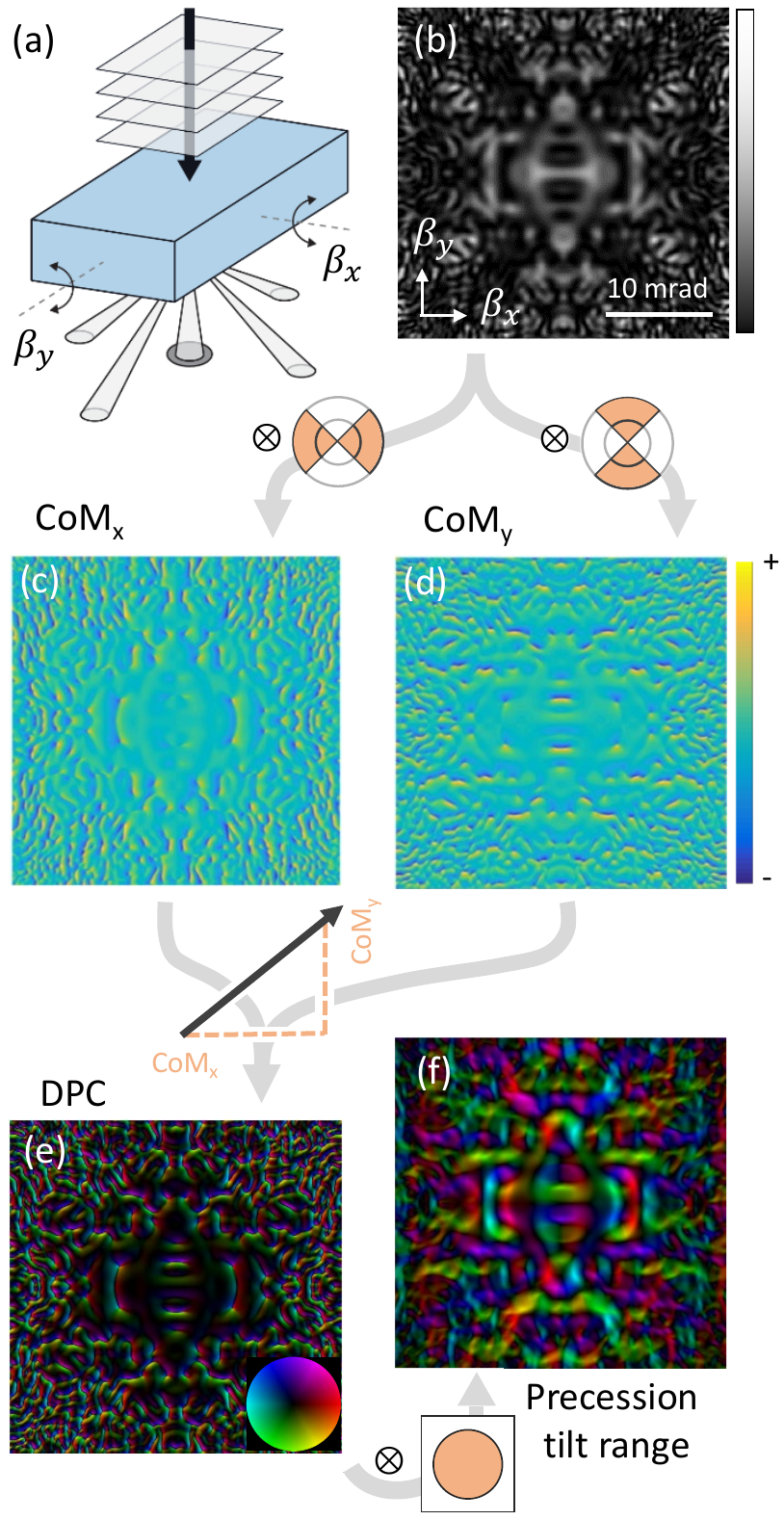}
	\caption{(a) Schematic of an incident plane wave being Bragg scattered through a crystal. The forward scattered signal---i.e. that recorded in a small on-axis disk detector---as a function of sample tilt angles $\beta_x$ and $\beta_y$ produces the so-called channelling map in (b). For any circular region of (angular) radius $\alpha$ less than half a Bragg angle, the intensity in (b) will also be the bright field intensity in STEM for a probe with probe-forming aperture semiangle $\alpha$. Consequently, convolving (b) with the detector response maps appropriate to calculating the approximate centre of mass (CoM) in the $x$- and $y$-direction produces the maps in (c) and (d), respectively, which show the CoM value produced by dynamical diffraction from the atomic electric fields for the relative beam-sample orientation $(\beta_x,\beta_y)$. (e) The CoM vector map with colour denoting direction and value (intensity) denoting amplitude. Precession amounts to an average of such maps over a range of precession angles. For a disk tilt pattern of precession angles, this can be achieved by convolving with a disk-shape binary mask. Applied to the map in (e) this produces the map in (f).}
	\label{F2}
\end{figure}

Though contributions to beam deflection from long-range electric fields and from dynamical diffraction from the atomic electric fields are not strictly additive \cite{taplin2016low}, the exploratory simulations in this manuscript consider only dynamical diffraction from atomic electric fields. This is because simulations incorporating both the atomic electric fields and the much-longer-range electric field of the p-n junction in Fig.~\ref{F1}(d)  are challenging: sampling finely enough to describe atomic scattering potentials across a field of view large enough to encompass structure like the p-n junction makes heavy demands on computer memory and calculation time. Considering only dynamical diffraction from atomic electric fields suffices to determine the scale of its contribution to the DPC signal and explore the extent to which precession may suppress it, while simplifying the calculation sufficiently that we can explore a much wider range of relevant parameters (such sample orientation and probe-forming aperture size). Our approach, summarised in Fig.~\ref{F2}, is as follows.

Consider a plane wave incident upon a crystalline sample. This results in Bragg scattering to discrete, narrow spots in the diffraction plane, as depicted in Fig.~\ref{F2}(a). Recording just the intensity of the central (``forward-scattered'') beam as a function of sample orientation parameterised by tilts $\beta_x$ and $\beta_y$ produces what we shall refer to as a channelling map,\footnote{It can also be called a channelling pattern, a two-dimensional rocking curve or a large-angle convergent-beam electron diffraction pattern \cite{reimer1997transmission}.} as shown in Fig.~\ref{F2}(b). In conventional TEM, such channelling maps are characteristic of the crystalline structure or, for spectroscopic signals, characteristic of the site location of various elements \cite{spence1983alchemi,rossouw1996statistical}. With a view towards DPC-STEM imaging, however, we give these channelling maps a slightly different interpretation. The intensity within any circular region of angular radius $\alpha$ on such a map can, provided the region's diameter is less than the smallest Bragg angle, be regarded as the intensity within the bright field disk when a STEM probe with probe-forming aperture semiangle $\alpha$ is incident upon the crystal with mistilt (relative to the optical axis) equal to the central point of the circle \cite{SFSMSKOMI1,wu2017correlative,haas2019direct}. Given the structure evident in Fig.~\ref{F2}(b), for all but the smallest of probe-forming apertures the intensity in the bright field disk is likely to be non-uniform. Therefore, the first moment of the diffraction pattern intensity within the bright field disk may be non-zero, even though there is no long range field present. If this contribution to the first moment signal varies across the field of view scanned, dynamical diffraction artefacts---by which, to reiterate, we mean beam deflection arising from scattering from the atomic electric fields (cf. long-range electromagnetic fields)---will be present in the DPC-STEM image. Variation across the field of view could arise from variation in local crystal orientation (equivalent to recentring the bright field disk to a different point in the channelling map) or variation in thickness (which changes the structure in the channelling map, as shown in section \ref{sec3}).\footnote{Further possibilities not captured by our analysis include the sample orientation varying with depth into the sample, and the local sample orientation and/or thickness varying on a scale smaller than the width of the probe. The latter is the likely cause of the DPC-STEM contrast in the localised surface-textural features in Fig.~\ref{F1}(d).}

Convolving the channelling maps with a response function for a segmented detector (assumed to have radius equal to the probe-forming aperture semiangle $\alpha$) therefore produces a map of the measured response on that detector as a function of sample orientation beneath the probe. This is shown in Figs.~\ref{F2}(c) and (d) for the centre-of-mass (CoM) in the $x$- and $y$-directions, and combined into a vector colour map in Fig.~\ref{F2}(e). Screw-type and edge-type topological defects \cite{SethnaBook} are again evident, in Figs.~\ref{F2}(c) and (d).  We will refer to maps like that in Fig.~\ref{F2}(e) as DPC-STEM channelling maps. The effect of precession---the averaging over a range of sample orientations---on such maps is obtained by convolving the maps with a binary mask defining the range of precession tilts. Figure \ref{F2}(f) shows the result of precession applied to Fig.~\ref{F2}(e).

\begin{figure*}[htb!]
	\centering
    \includegraphics[scale=1.7]{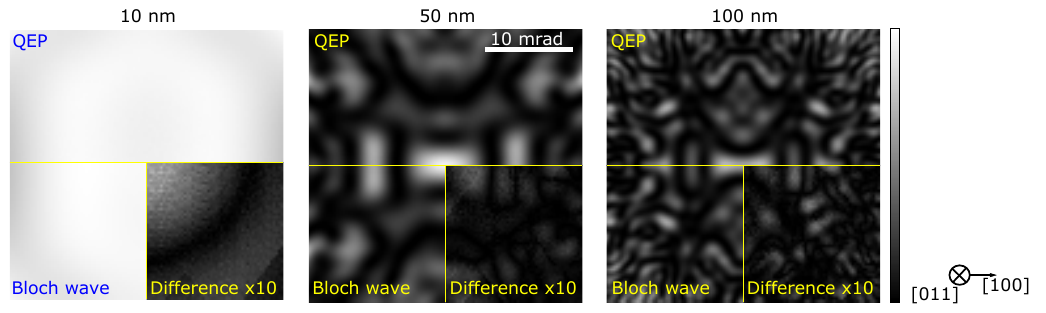}
	\caption{Channelling maps for three different thicknesses, comparing the results of calculations using the QEP model, the Bloch wave model and their difference (essentially the intensity due to thermally scattered electrons). Results for each thickness are shown minimum-black to maximum-white, with the QEP and Bloch wave panels on the same colour scale while their difference is scaled up by a factor of 10. The differences are small and show that the qualitative feature detail is adequately captured by the Bloch wave model.}
	\label{F3}
\end{figure*}

Comment is warranted on the method for simulating channelling maps like that in Fig. \ref{F2}(b). Of the common approaches to electron scattering simulation, the most accurate is generally considered to be the frozen phonon \cite{loane1991thermal} or quantum excitation of phonons (QEP) \cite{forbes2010quantum} model, usually implemented via the multislice method \cite{K8}. It accounts for both elastically and thermally scattered electrons by averaging the measured signal (here the forward-scattered intensity) calculated from many different configurations of atoms consistent with atomic thermal motion in the crystal. However, the large number of configurations needed to produce converged calculations and the linear scaling of the calculations with sample thickness is prohibitive for systematic exploration of fine features in diffraction patterns from very thick samples. Channelling maps have more usually been calculated using the Bloch wave method \cite{spencer1972dynamical,rossouw1991treatment,josefsson1994k}. For diffraction intensity, this approach should correctly describe the elastic scattering, including the reduction in elastic component due to thermal scattering, but it omits the intensity contribution from thermally scattered electrons. Further, in the zero order Laue zone approximation (a high energy approximation which neglects the curvature of the Ewald sphere), periodicity relations among the Bloch wave coefficients \cite{kastner1993many} mean that distinct calculations only need to be performed for the different $(\beta_x,\beta_y)$ orientations within the first Brillouin zone. Figure \ref{F3} compares channelling maps calculated via QEP and Bloch wave methods (with the approximations stated above) for three different thicknesses. The differences are quantitatively small, and do not produce an appreciable change in qualitative features of the channelling maps. As such, the Bloch wave approach has been used in all simulations that follow. Throughout this paper, all simulations assume 300 keV electrons and a GaAs sample aligned close to the [011] zone axis orientation. If not otherwise specified, the sample thickness is 200 nm, comparable to the sample thickness in Fig.~\ref{F1}(d).

\section{Dependence of channelling map complexity on thickness}
\label{sec3}

\begin{figure*}[htb!]
	\centering
		\includegraphics[scale=0.9]{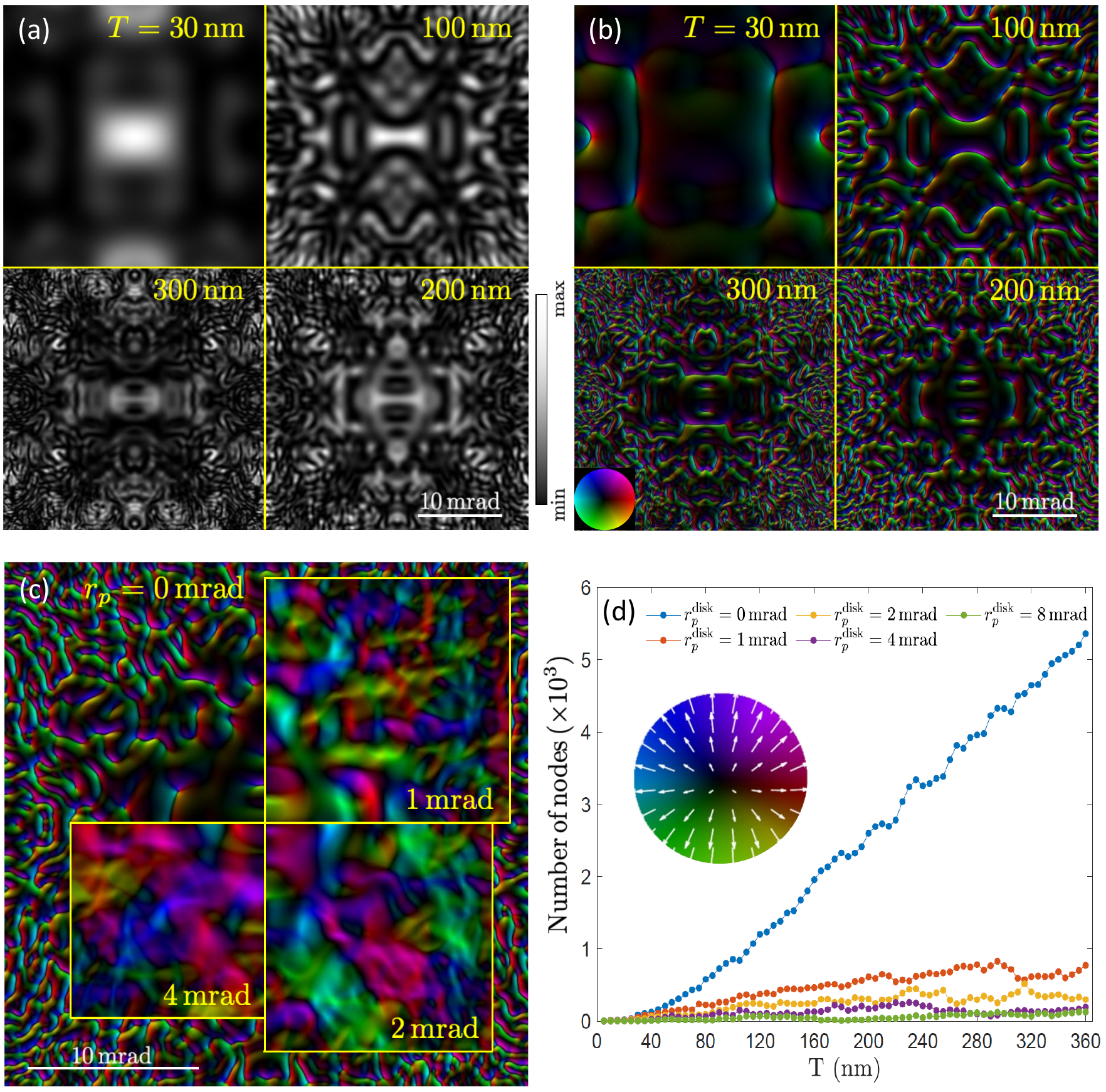}
	\caption{(a) Channelling maps for four different thicknesses of GaAs. The centre of the map corresponds to the exact [011] zone axis orientation. The accelerating voltage assumed is 300 kV. (b) DPC-STEM channelling maps for the same sample and thicknesses assuming a STEM probe with probe-forming aperture semiangle $\alpha = 0.2$ mrad and an eight-segment detector (two rings of equal annular width, each divided into quadrants). (c) The 200 nm map from (b) with insets corresponding to disk-shaped precession ranges with radius $r_p$ as shown. (d) Number of nodes for a span of thicknesses and precession ranges. An example node is provided in the inset, with sample vectors showing the associated texture.}
	\label{F4}
\end{figure*}

Figure \ref{F4}(a) shows channelling maps for four different thicknesses, and Fig.~\ref{F4}(b) shows the corresponding DPC-STEM channelling maps. Before discussing the thickness dependence, it is worth elaborating upon the connection between the appearance of these two kinds of images. The most prominent structure in the DPC-STEM channelling maps (Fig.~\ref{F4}(b)) is the filaments of coloured contrast, usually with two sides of diametrically-opposed colour (see inset colour wheel) separated by a fine line of low contrast. This structure can be understood by appreciating that, for all patches in the channelling maps with features notably broader than the width of the probe, each point in the DPC-STEM channelling map essentially reflects the local gradient at the corresponding point in the channelling map. Large features in the channelling maps imply small local gradients, and so correspond to extended dark regions in the DPC-STEM channelling maps. The coloured filaments in the DPC-STEM channelling maps correspond, in topographic terms, to sharp ridges and valleys in the channelling maps, since the local gradient changes rapidly in direction across such features.

The amount of structure present in the maps in Figs.~\ref{F4}(a) and (b), or more specifically the rapidity of variation of the signal with variation in $(\beta_x,\beta_y)$, is seen to increase with increasing thicknesses. This implies that the thicker the sample the less variation in local sample orientation is needed across the STEM probe scan to produce notable variation in DPC-STEM signal via dynamical diffraction.\footnote{This interpretation assumes $\alpha$ to be quite small on the scale of these images, which is indeed typically the case for imaging long range fields---for the data in Fig.~\ref{F1}, $\alpha \approx 0.1$ mrad. For much larger apertures, the trend might reverse if increasingly fine structure of the channelling map within the aperture was thereby averaged over.} Figures \ref{F4}(a) and (b) also show that the rapidity of variation with $(\beta_x,\beta_y)$ for a given thickness broadly increases for orientations further away from the central zone axis orientation. Being somewhat off the zone axis may thus actually increase the amount of dynamical-diffraction-induced variation in DPC-STEM images due to variation in local sample orientation across the scan, relative to what might be obtained closer to the zone axis.

From these maps we can also grasp the qualitative mechanism by which precession helps reduce the impact of dynamical diffraction on DPC-STEM imaging: averaging over orientations suppresses these rapid variations to some extent, and reduces the sensitivity of the signal to small changes in local sample orientation. Figure \ref{F4}(c) reproduces the DPC-STEM channelling map for the 2000 {\AA} case with overlays showing the same map after precession assuming a solid disk tilt angle pattern (with radius as shown). The overlays are positioned in correct accordance with the $(\beta_x,\beta_y)$ coordinates of the underlying map, taking advantage of the top-bottom and left-right symmetry to facilitate visual comparison between the overlaid maps. As the precession radius gets larger, the features in the DPC-STEM channelling maps become increasingly spread out, meaning that a larger variation in local sample orientation across the scan would be needed to produce comparable variation in DPC-STEM signal via dynamical diffraction.

\begin{figure*}[htb!]
	\centering
		\includegraphics[scale=1.0]{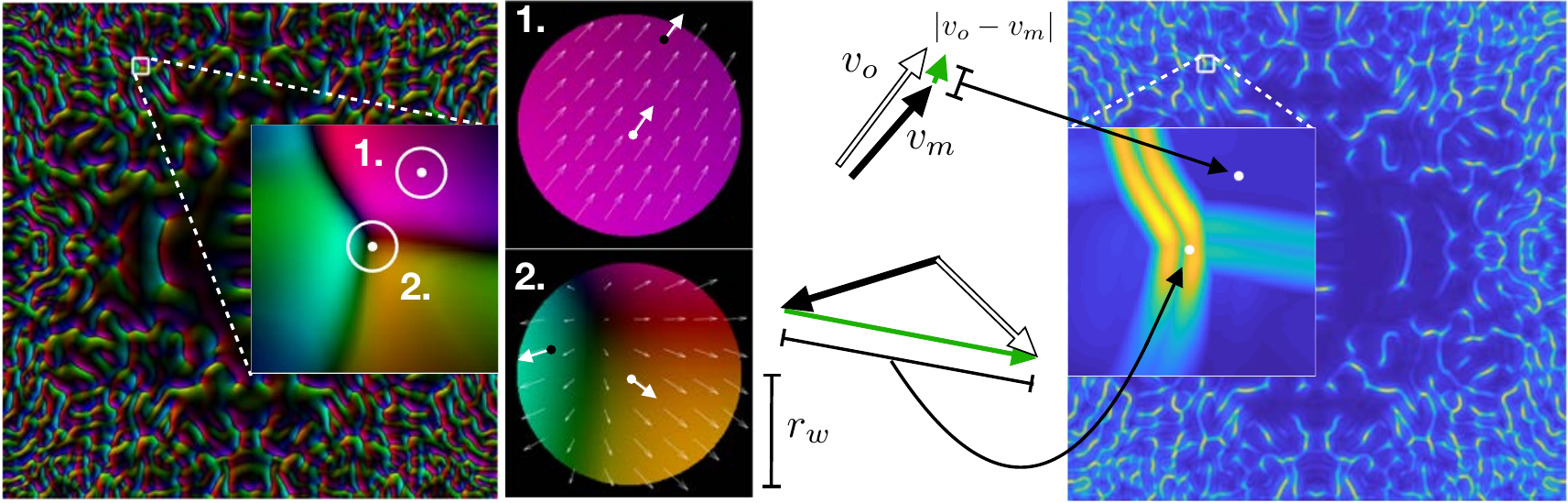}
	\caption{Schematic of the construction of the (dynamical diffraction) artefact metric. In the magnified portion of the DPC-STEM channelling map on the left, two windows are identified (white circles with central white dots). Enlarged versions of these windows are shown in the centre-left panel with CoM quiver plots overlaid. In the upper quiver plot, all arrows have similar magnitude and direction: while dynamical diffraction is giving a contribution to the DPC signal, it varies minimally across the different local sample orientations within this window. In the lower quiver plot, the arrows vary appreciably in both magnitude and direction: not only does dynamical diffraction give an appreciable contribution to the DPC signal, but it varies appreciably across the different local sample orientations within this window. To quantify this, we define an artefact metric---describing the beam deflection arising from scattering from the atomic electric fields (cf. long-range electromagnetic fields)---equal to the magnitude of the maximum \emph{difference} vector between the CoM vector at the central point of the window and at all other points in the window. This construction is shown in the centre-right panel for the two points at the centre of our example windows, and then as a map over the full $(\beta_x,\beta_y)$ range on the right.}
	\label{F5}
\end{figure*}

The quantitative consequences of both dynamical diffraction in the sample and precession as an imaging strategy will be explored in the following section, but it is informative at this point to introduce a measure of the structural complexity of the maps in Fig.~\ref{F4}(a)-(c). Figure \ref{F4}(d) plots, as a function of sample thickness, the number of nodes or ``defects'' in the vector maps, points  about which centre of mass shifts either radiate out from or in to. (These points are approximately congruent to local maxima or minima in the channelling map.) Though not strictly monotonic, without precession the number of these features increases approximately linearly with sample thickness. For samples thicker than about 50 nm, the number of these features is significantly suppressed via a disk tilt map of precession with semiangle 1 mrad. Precession through larger angles provides further improvements, though to a proportionally smaller (sub-linear) extent. Since the exact number of defects depends on the field of view, the significance of this plot lies in the trends it shows. Moreover, while these defects are largely unambiguous to identify, their presence is not of itself a measure of the potential error in DPC signal due to dynamical diffraction. For that, we need a different metric.

\section{Parameter exploration of the impact of dynamical diffraction on DPC signal}

The DPC-STEM signal that results from dynamical diffraction for a given orientation of the crystal relative to the optical axis of the STEM probe is given by a single point in the DPC-STEM channelling map. An expanded portion of a DPC-STEM channelling map is shown on the left in Fig.~\ref{F5}. The white dots (at the centre of the white circles---more on those below) indicate two particular orientations of the sample. The DPC-STEM signal from dynamical diffraction being non-zero at such points means that the first moment of the bright-field disk with the sample present is different from what it would be if the sample were not present. Nevertheless, if the sample orientation was the same across the entire region over which the STEM probe is scanned, the contribution from dynamical diffraction would be constant and so to a good approximation would not adversely affect the mapping of long-range electromagnetic fields.\footnote{Taplin \emph{et al.} \cite{taplin2016low} explored the combination of constant long-range electric fields and dynamical diffraction, and showed that the effect of dynamical diffraction on the first moment signal was not quite linear with field strength, which could produce a correction to a long-range electromagnetic field map even if the atomistic structure were uniform.}

\begin{figure}[htb!]
	\centering
		\includegraphics[scale=1.0]{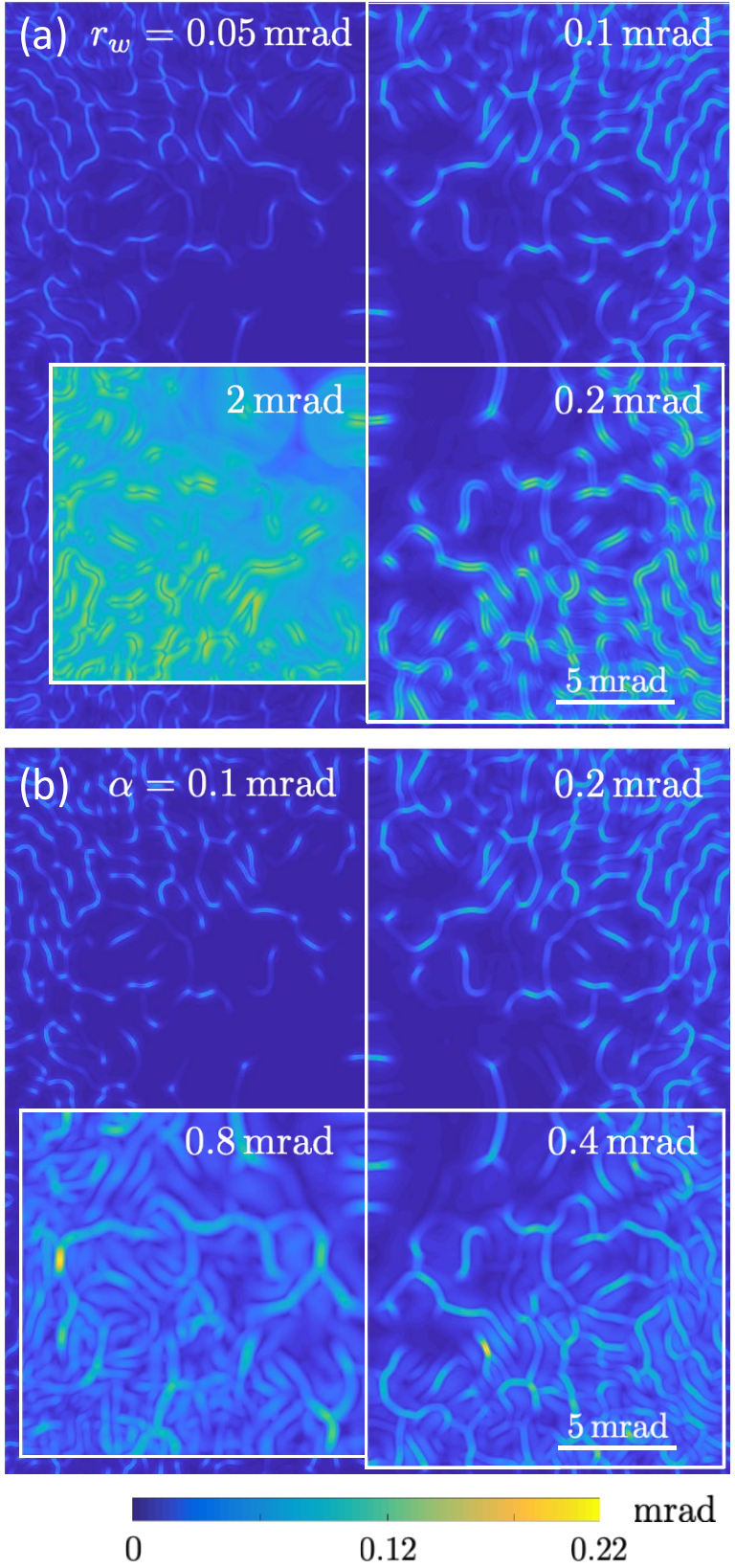}
	\caption{Artefact maps for (a) various window radii at fixed probe-forming aperture semiangle $\alpha=0.2$ mrad, and (b) various probing-forming aperture semiangles at fixed window radius $r_w=0.1$ mrad. All maps are on the same scale. Increasing window radius tends to increase the magnitude of the artefact metric, and the extent of $(\beta_x,\beta_y)$ points over which it is appreciable. Increasing aperture size produces a qualitatively similar trend but to a lesser degree.
	}
	\label{F6}
\end{figure}

More likely, due to sample bending, the local sample orientation will vary across the STEM probe scan region. Expecting the change in local sample orientation to be smooth, the range of orientations present defines some continuous window of the DPC-STEM channelling map. The map on the left of Fig.~\ref{F5} uses white circles to highlight two such windows---chosen as disks for simplicity, though the distribution of orientations present in a given field of view of a real sample is unlikely to be so symmetric. These windows (with radius $r_w$) are magnified in the centre-left part of Fig.~\ref{F5}. The lower window displays considerable variation in the CoM vector, meaning that dynamical diffraction from atomic electric fields would produce a notable variation in the DPC-STEM signal across the field of view (irrespective of whether or not long-range electromagnetic fields are present). The upper window has smaller variation in the CoM vector, meaning that dynamical diffraction artefacts would be correspondingly smaller.

For a more quantitative exploration of the impact of dynamical diffraction artefacts on DPC-STEM imaging, we define a metric of the variability in DPC signal due to changes in local sample orientation to be the maximum magnitude of the difference vector between the CoM vector at the centre of a disk-shaped window and all of the other CoM vectors within the window. This is depicted in the central part of Fig.~\ref{F5}. (Minor variants on the chosen metric, like the mean difference in magnitude, were also explored but showed similar trends.) This variability metric, which may be viewed as an order-parameter field that serves to highlight the topological defects mentioned earlier in the paper \cite{SethnaBook}, can be plotted as a channelling-type map as shown on the right in Fig.~\ref{F5}. These maps provide a quantitative measure of the variation in the DPC-STEM channelling map in the vicinity of each orientation, and in what follows we shall refer to them as artefact maps. Recalling from our discussion of Fig.~\ref{F4}(b) that the most prominent structure in the DPC-STEM channelling maps are filaments of coloured contrast, maxima in the artefact maps tend to occur along those filaments since these are the regions where the magnitude of the DPC signal is large and its direction is rapidly changing, meaning that the DPC signal changes significantly for small variation in local sample orientation.

For a probe-forming aperture semiangle $\alpha=0.2$ mrad, Fig.~\ref{F6}(a) again takes advantage of the near centrosymmetry of the channelling maps to compare the artefact maps for four different window radii. A small window radius corresponds to very little variation in local sample orientation across the field of view; a larger window radius corresponds to a larger variation. Intuitively, increasing the range of sample orientations present across the field of view might be expected to increase the variability of the DPC signal across that field of view and so the magnitude of the artefact map. Figure \ref{F6}(a) shows this is true only to a point: increasing the windows from 0.05 mrad to 0.2 mrad approximately doubles the maximum deflection, as evident in the increased contrast of the filamentary features. However, as the window is further increased the maximum deflection starts to saturate. Essentially, there is an approximate upper limit on the magnitude of CoM shifts possible for intensity within the bright field disk based on the scale of features in the channelling map,\footnote{Since we are calculating the CoM of the bright field region, the absolute upper limit is the radius of the bright field disk.} and, as per the filamentary structure of the DPC-STEM channelling maps, sizeable oppositely-directed CoM vectors tend to occur in close proximity to one another. Once the window is wide enough to encompass these filaments, further widening of the window will not further increase the maximum deflection within the window. That said, widening the window does increase the proportion of angles in the artefact map for which appreciable variation occurs somewhere within the window, as evident in the increased width of the filamentary features with increasing window size in Fig.~\ref{F6}(a).

Figure \ref{F6}(b) shows a similar visualisation, now keeping the window radius fixed at $r_w = 0.1$ mrad and comparing the maps for different probe-forming aperture semiangles $\alpha$. Whereas increasing the window amounts to changing the size of the region over which the variation is measured within a given DPC-STEM channelling map, increasing $\alpha$ changes the DPC-STEM channelling map itself. Nevertheless, because the aperture sizes considered here are still reasonably small on the scale over which the main structural features in the channelling maps vary, the trend is quite similar though less pronounced: the filaments get wider and their intensities increase.

The variation in local sample orientations (as characterised by the window size) is, of course, not controllable in the way that aperture semiangle is, but these artefact maps show that both smaller apertures\footnote{Smaller apertures are anyway expected to improve field mapping sensitivity, since a given deflection (determined by the field strength) may more reliably be measured if it is a larger fraction of the bright field disk size \cite{haas2019direct}. While Fig. \ref{F6}(b) suggests that the artefact metric magnitude may become a larger \emph{fraction} of the bright field disk size for smaller probe-forming aperture angles, the more important comparison is with the size of the deflection due to the field strength of interest, and for this purpose a smaller absolute value of the artefact metric is deemed better.} and smaller local sample orientation variation are preferable, since both increase the range of average orientations for which the maximum deflection is particularly low and reduce the scale of the artefact metric for a given average orientation. But perhaps the most significant point about Fig.~\ref{F6} is its scale: for the less favourable average orientations, the artefact metric is a sizeable fraction of the bright field disk radius $\alpha$. This is consistent with the experimental results of Fig.~\ref{F1}(b), where the contrast attributed to variation due to dynamical diffraction contrast dominates that due to the p-n junction.

\begin{figure*}[htb!]
	\centering
		\includegraphics[scale=1.0]{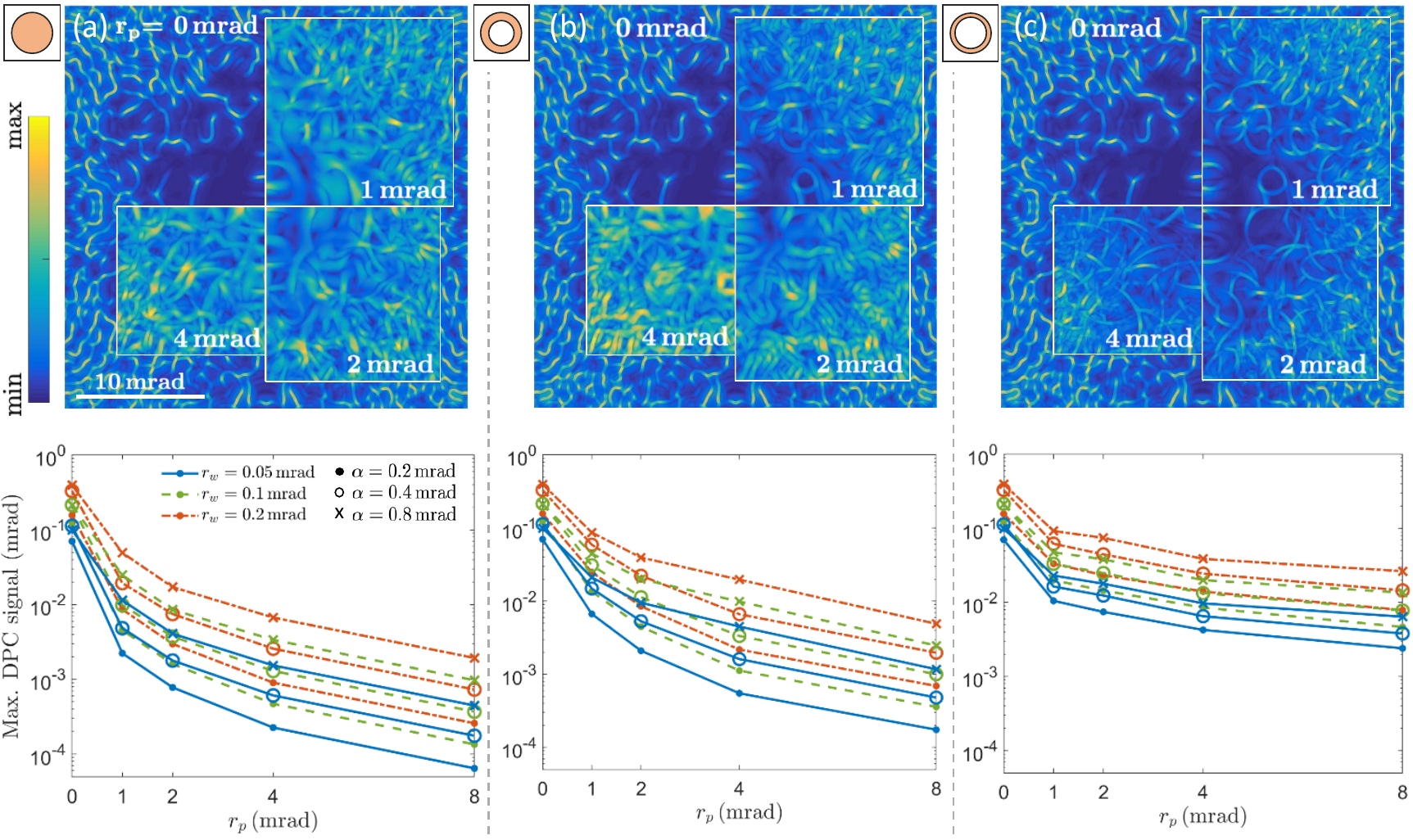}
	\caption{For a 200 nm thick GaAs sample around the [011] zone axis orientation, a probe-forming aperture semiangle $\alpha = 0.2$ mrad, a window size of $r_w = 0.1$ mrad, and for precession tilt angle patterns of (a) a disk, (b) a broad annulus, and (c) a narrow annulus, the upper half of the figure shows artefact maps for a variety of precession ranges. Each map is plotted on its own colorbar from its maximum to minimum range. The plots in the lower half of the figure show the maximum values of each artefact map, as well as those for two further probe-forming aperture semiangles (0.4 and 0.8 mrad) and two further sample orientation variation windows ($r_w = 0.05$ and $0.2$ mrad).}
	\label{F7}
\end{figure*}

\section{Precession}
\label{sec:precession}

The previous section shows that using judicious choice of sample orientation to mitigate dynamical diffraction artefacts from the variation in local sample orientation across the field of view not only requires care but indeed will only be possible in some circumstances, specifically, when there is modest orientation variation across the field of view and a sufficiently small probe-forming aperture semiangle is used. In this section we quantify the extent to which precession can overcome these constraints.

Figure \ref{F7} explores the effect of precession on the artefact maps for our GaAs [011] case study. Because the effect of precession on the range of the artefact maps is quite pronounced, we separate out the qualitative features from the quantitative range information. Assuming a 200 nm GaAs sample and a probe-forming aperture semiangle $\alpha=0.2$ mrad, in the upper row each map and their overlays are individually normalised such that the colour scale matches their full minimum-to-maximum range. The quantitative maximum value in each artefact map is shown (on a logarithmic scale) in the plots in the lower row, together with the same quantity for other values of probe-forming aperture semiangle $\alpha$ and degrees of local sample orientation variation described by window radius $r_w$.

Figure \ref{F7}(a) considers a solid disk tilt angle pattern for precession. The effect of precession can, as per the discussion of Fig.~\ref{F2}(f), be described as a convolution of the DPC-STEM channelling maps with the precession tilt angle pattern. While the construction of the artefact maps is such that the precession convolution and this construction do not strictly commute, nevertheless the artefact maps in Fig.~\ref{F7}(a) show that increasing the range of precession (the radius of the disk tilt angle pattern) essentially broadens the features in the maps while reducing the contrast, i.e. the variation from point to point. This should help by reducing the need to carefully select a particular orientation that is less sensitive to channelling effects. But the main advantage is seen in the quantitative information in the plot: a precession range of 1 mrad reduces the artefact metric by around an order of magnitude. While the additional advantage gets smaller with subsequent doubling of the precession range, for a precession range of 8 mrad the reduction in artefact metric is over two orders of magnitude. This precession range is comparable to that used in Fig.~\ref{F1}(d) (although the precession pattern there was square rather than circular), and the reduced dynamical diffraction artefacts of that experimental, precessed DPC-STEM image are broadly consistent with the simulations of Fig. \ref{F7}(a).

Since precession can be challenging to implement mechanically, it is worth asking whether the full range of orientations in the disk tilt angle pattern are really necessary, or whether similar advantage might not be conferred with an annular pattern. This is explored in Figs.~\ref{F7}(b) and (c), for a broad annulus and narrow annulus precession pattern, respectively (in each case $r_p$, the precession range, is defined to be the outer edge of the annulus). These results show that the disk precession pattern confers more advantage than annular precession patterns. While the initial advantage of some precession over no precession is not so different for the three precession patterns, being a reduction of about an order of magnitude, the advantage for further increasing the precession range is minimal for the narrow annulus precession pattern but significant for the disk pattern.

\section{Thickness variation}

\begin{figure*}[htb!]
	\centering
		\includegraphics[scale=1.0]{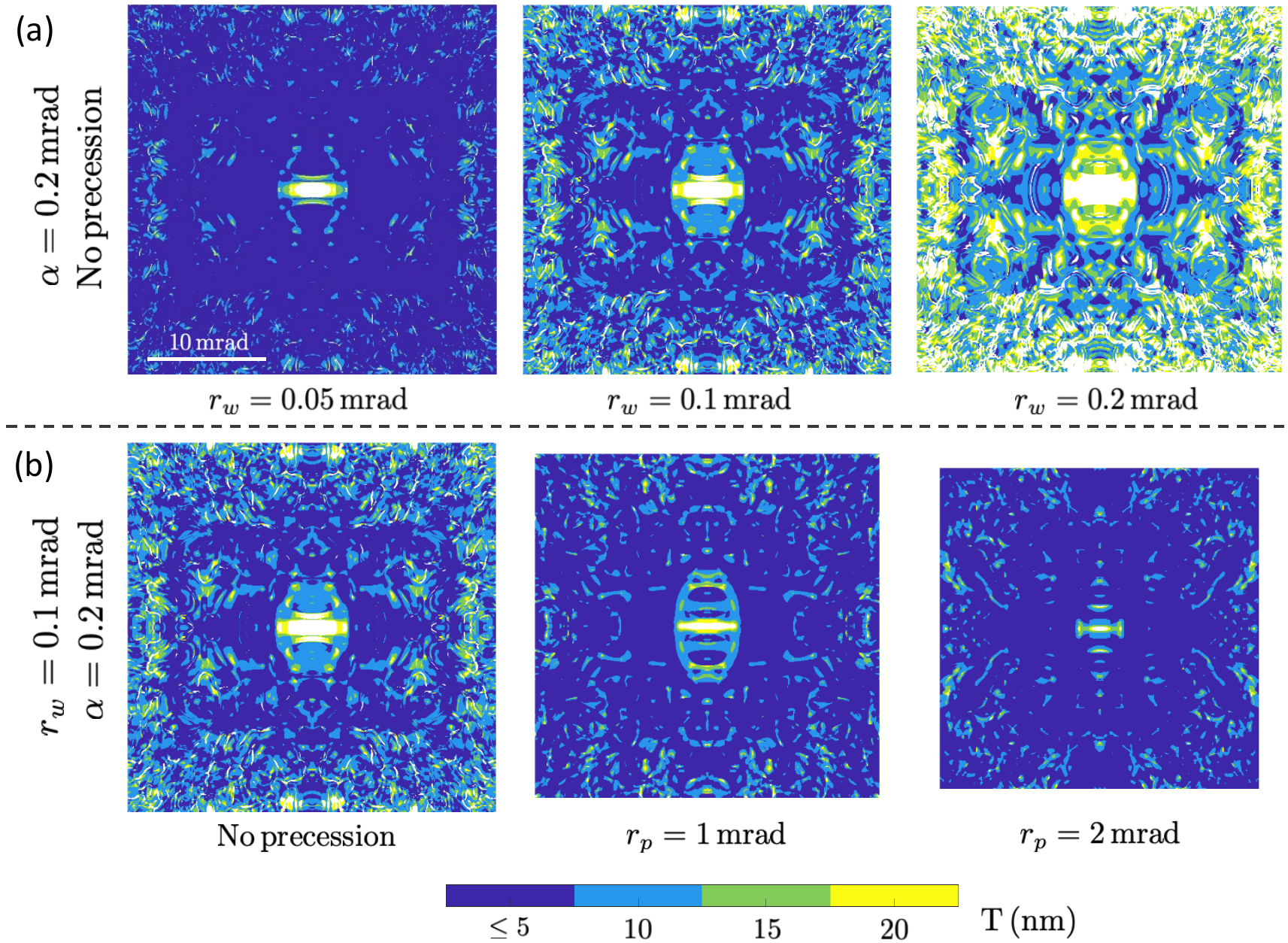}
	\caption{The thickness variation required to obtain the same artefact metric as that due to local sample orientation variation within a given sample orientation variation window of radius $r_w$. A discrete colour scale is used such that, for instance, regions coloured blue-green indicate that the same artefact metric as that from the variation of local sample orientation within a given window size would be realised for a sample thickness variation between 5 and 10 nm. The maximum thickness variation considered is $20$ nm, equivalent to ten percent of the total sample thickness. Regions coloured white require a still greater thickness variation to achieve the same artefact metric as that from the variation of local sample orientation within a given window size. (a) Varying window size for a fixed probe-forming aperture semiangle $\alpha = 0.2$ mrad and no precession. (b) Varying precession for a fixed window size of $r_w = 0.1$ mrad and probe-forming aperture semiangle of $\alpha = 0.2$ mrad. }
	\label{F8}
\end{figure*}

So far our analysis has concentrated solely on the dynamical diffraction artefacts arising from variations in local sample orientation. However, as per the cartoon in Fig.~\ref{F1}(a), sample thickness variation across the probe scan region could also produce dynamical diffraction artefacts in the DPC-STEM signal. Figures \ref{F3} and \ref{F4} show that for significantly different thicknesses the channelling maps are themselves significantly different. On grounds of continuity, bolstered by the indirect evidence of Fig.~\ref{F4}(d), we expect these maps to vary smoothly with thickness. Since variation in local sample orientation and local sample thickness could well occur together, we explore the scale of the effect of the latter by comparing it to the former. For various combinations of probe-forming aperture semiangle, sample orientation variation window and precession radius, the maps in Fig.~\ref{F8} show the sensitivity of DPC signals to variations in the sample thickness in the form of the variation in thickness (away from the 200 nm reference thickness) required at each $(\beta_x,\beta_y)$ point to give the same artefact metric as that from the variation of local sample orientation within a given window size $r_w$.

Figure \ref{F8}(a) shows maps for three different window sizes, assuming a fixed probe-forming aperture semiangle of 0.2 mrad and no precession. Greater window sizes permit greater variations in the local sample orientation and so larger artefact metrics. As a consequence, for the larger window size and for most orientations shown, the thickness variation would need to be around 15-20 nm---almost 10\% of the total sample thickness---for thickness variation to contribute as much to dynamical diffraction artefacts as local sample orientation variation. Since variations on this scale are expected to be perceptible in high-angle annular dark field images, their absence suggests that sample orientation variation is the dominant mechanism producing dynamical diffraction artefacts. Conversely, if the variation in local sample orientation over the field of view was the more modest 0.05 mrad of the left-most panel in Fig.~\ref{F8}(a) then thickness variations less than 5 nm would produce dynamical diffraction artefacts comparable to those from local sample orientation variation (though note that, as per Fig. \ref{F6}(a), the artefact metric is quite small near the exact [011] zone axis orienatation).

Figure \ref{F8}(b) shows maps for a fixed probe-forming aperture semiangle of 0.2 mrad and a local sample orientation variation window of 0.1 mrad but for differing degrees of precession. As per the previous section, precession can significantly suppress the dynamical diffraction artefacts from local sample orientation variation, being effectively an average over such variations. For increasing amounts of precession, Fig.~\ref{F8}(b) shows that the minimum thickness variation producing the same artefact metric gets successively smaller: precession is effective at suppressing dynamical diffraction artefacts from local sample orientation variation but notably less successful at suppressing those due to thickness variation.

\section{Comparison with experiment}

Though motivated by findings from earlier experimental work (see Fig.~\ref{F1} or, for further details, Ref. \cite{nakamura2017differential}), we have thus far presented an exploration of the effects of some key parameters---thickness, aperture size, sample orientation variation, precession range---through simulation. Let us round out the discussion with a comparison against experimental data.

\begin{figure*}[htb!]
	\centering
		\includegraphics[scale=1.0]{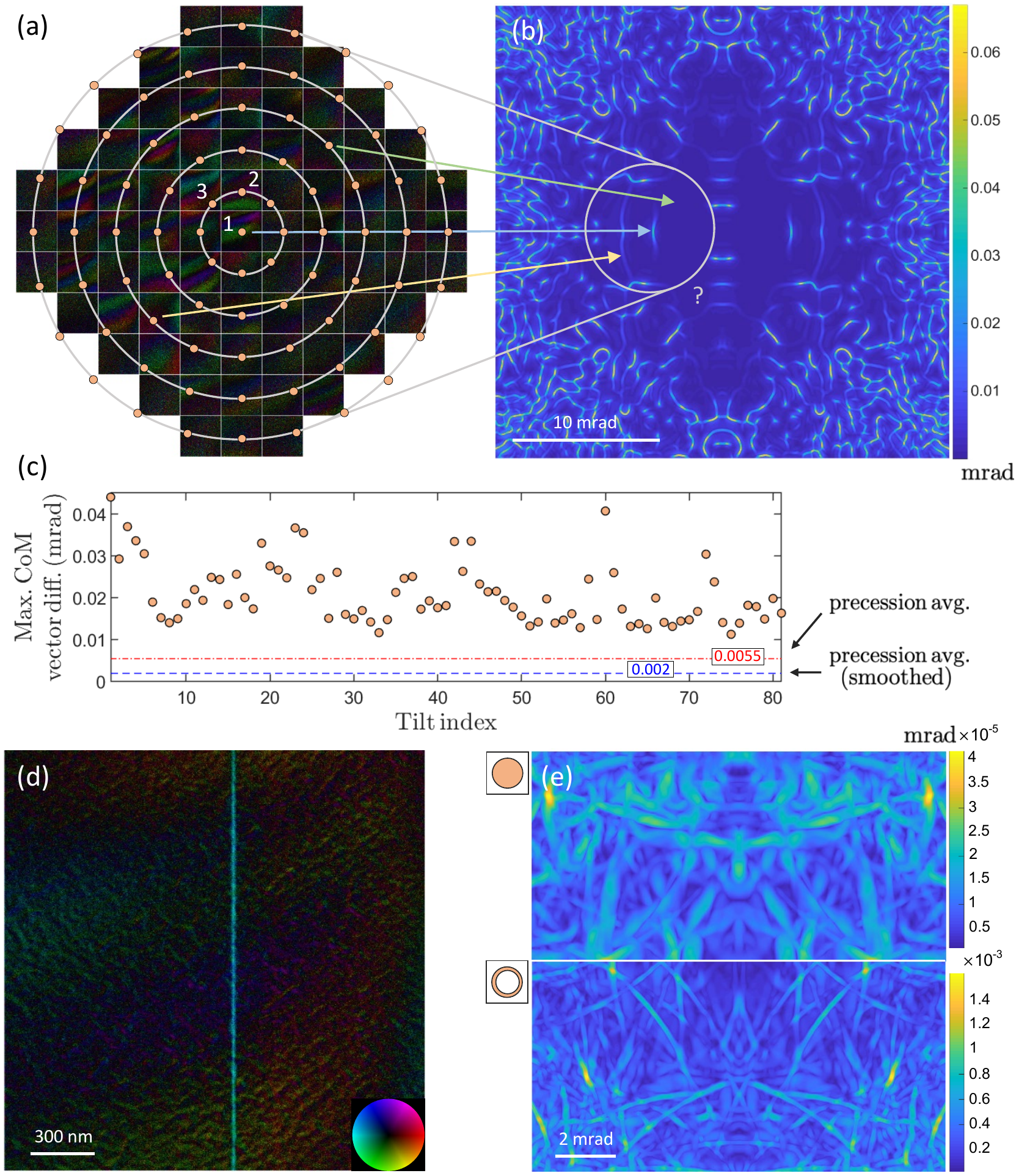}
	\caption{(a) Tableau of DPC-STEM images from a GaAs sample near the [011] zone axis orientation (from a different sample to that shown in Fig.~\ref{F1}). Overlaid is the precession tilt map, comprising a discrete set of points in nested circles, with the outermost having radius 0.5 degrees. (b) Artefact map for 240 nm thick GaAs [011] assuming a sample orientation variation window of radius $r_w=0.05$ mrad. Though the exact orientations at which the DPC-STEM images in (a) were recorded are unknown, overlaid on (b) is a candidate location based on the similarity between the regions in which dynamical diffraction artefacts would be large and those in which they would be small. (c) Plot of the maximum CoM vector difference in the individual tilt DPC-STEM images in (a), after binning down by a factor of 4 to reduce noise. The first three tilt index labels are shown in (a). Also plotted is the maximum CoM vector difference of the precession-averaged DPC-STEM image after binning down by a factor of 4 to reduce noise (red dot-dashed line) and after further smoothing to suppress some fine structural features that appear to be distinct from the larger-range CoM variation attributed to dynamical diffraction (blue long-dashed line). (d) Precession-averaged DPC-STEM image formed from the images in (a). (e) Artefact maps as per (b) but including precession over a solid disk (upper) and thin annulus (lower), both with outer radius 0.5 degrees.}
	\label{F9}
\end{figure*}

The junction field strength in the previous sample \cite{nakamura2017differential} showing signs of appreciable degradation under multiple exposures, a new p-n-junction-containing sample of GaAs with thickness 240 nm was prepared in [011] zone axis orientation by focused ion beam milling. STEM DPC imaging was carried out on a JEM-ARM300F operating at 300 keV with a probe-forming aperture semiangle of 129 $\mu$rad and using a SAAF-OCTA detector at camera length such that the bright field disk extended to the radial mid-point of the outer ring of the detector. Precession was undertaken by manually tilting the sample, with images being obtained at a range of tilts shown as dots in Fig.~\ref{F9}(a) that form a series of concentric rings with the radius of the outermost ring being 0.5 degrees. Figure~\ref{F9}(a) further shows a tableau of the CoM vector maps.

Figure~\ref{F9}(b) shows the artefact map simulated for the experimental configuration described above, assuming a sample orientation variation window of radius $r_w=0.05$ mrad. Though the exact orientation of the central tilt in Fig.~\ref{F9}(a) is not known, a to-scale copy of the precession tilt map has been overlaid on the artefact map. This location is plausible because it is broadly consistent with structure in the tableau of the DPC-STEM images. Some of the DPC-STEM images, reminiscent of Fig.~\ref{F1}(b), show appreciable contrast variation not only at the junction location but more widely across the field of view, while others show very little such contrast. The region marked on the artefact map likewise contains some areas where the artefact metric is relatively large (blue and orange arrows), indicating orientations around which small variations in local sample orientation across a field of view would produce large variations in DPC signal due to dynamical diffraction from the atomic potential in the absence of the long range field, and others where the artefact metric is small (green arrow), implying changes in dynamical diffraction due to local sample orientation variation would not appreciably alter the DPC signal.

To make this more quantitative, Fig.~\ref{F9}(c) plots an estimate of the maximum variation in DPC vector within each of the individual DPC-STEM images in the precession series (excising the region containing the p-n junction since that deflection is due to the long-range electric field of interest rather than being a dynamical diffraction artefact). We say ``estimate'' because the maximum in the raw data was sensitive to the noise level. To reduce this, the values in Fig.~\ref{F9}(c) are the maxima after binning the individual DPC-STEM images down by a factor of four. These values are not a perfect match for the range on the artefact map, perhaps suggesting that our (somewhat arbitrary) assumption of $r_w=0.05$ mrad radius sample orientation variation window may be larger than the actual orientation variation on the sample across this field of view. But given the orientation variation in practice is anyway unlikely to conform to a perfect disk, we deem the experiment and simulations encouragingly consistent.

Figure~\ref{F9}(d) shows the DPC-STEM image after precession averaging. Reminiscent of Fig.~\ref{F1}(d), precession averaging has suppressed the contrast across the field of view such that the p-n junction stands out clearly. Figure~\ref{F9}(e) shows artefact maps assuming a solid disk precession pattern (upper) and thin annulus (lower) with outer radius 0.5 degrees. The scales on their respective colour bars shows that, consistent with section \ref{sec:precession}, the prediction for disk precession is a suppression of dynamical diffraction artefact size by almost three orders of magnitude while the prediction for the thin annulus is a suppression of dynamical diffraction artefact size by one-to-two orders of magnitude. The horizontal lines in Fig.~\ref{F9}(c) seek to convey the scale of the contrast (away from the junction) remaining after precession. The upper straight-horizontal line was obtained using identical processing to that applied to the individual DPC-STEM images in the precession series, but shows only a relatively modest reduction, much less than the simulated predictions. Close inspection of the precession-averaged map in Fig.~\ref{F9}(d) shows part of the reason: fine features of appreciable contrast are visible that are spatially large enough that the noise-suppression averaging does not remove them but notably finer than the longer range variation attributed to changes in dynamical diffraction condition. Close inspection of the individual tilt images shows that they too have similar-sized features (though in that case they are smaller than the larger-scale contrast variation in dynamical-diffraction-sensitive orientations). We tentatively attribute these features, which have length scale comparable to the probe width, to surface roughness. The lower horizontal line in Fig.~\ref{F9}(c) is an estimate of the scale of the longer range variation attributed to changes in dynamical diffraction condition after further averaging to suppress the finer structures. This shows dynamical diffraction contrast to have been suppressed by about an order of magnitude. This is still less than the annular precession prediction, but bear in mind that the artefact maps do not include any contribution from thickness variation and that the discrete range of tilts in the precession mesh here implies less averaging than that in the simulations assuming a continuous annulus or disk. It thus seems that the extent of suppression of dynamical diffraction artefacts obtained experimentally may depend strongly on the extent and smoothness of the precession scan region.

\section{Conclusion}

Through simulation, we have explored the contribution of dynamical diffraction to the variation of DPC-STEM signal across a scan region within which there can be variation in local sample orientation (due to specimen bending) and in thickness. Even relatively modest variations in these quantities can produce variations in the DPC-STEM signal that are a significant fraction of the bright-field disk radius. Precession over a disk-shaped tilt angle pattern a few milliradian in radius can suppress the dynamical diffraction artefacts due to local sample orientation variation. Simulations predict that this suppression could be a few orders of magnitude, but experimental results give a more modest single-order-of-magnitude suppression of dynamical diffraction artefacts, though perhaps only because the discrete set of points in the precession tilt map does not provide as much averaging as a continuous range of precession angles. Nevertheless, our findings reinforce that such precession is an effective strategy for improving the reliability of long-range electromagnetic field mapping.

The results presented have all assumed a GaAs sample in the close vicinity of the [011] zone axis. As noted by Haas \emph{et al.} \cite{haas2019direct}, the structure of the channelling map can differ notably for different samples and orientations. The utility of the present work to other cases lies not so much in the specific numerical values but rather in the trends (with thickness, aperture size, precession range, etc.) and also in the general strategy, which can be applied to any sample of known crystallographic structure.

\section*{Acknowledgements}

This research was supported by the Australian Research Council Discovery Projects funding scheme (Project DP160102338). This work was supported by the JST SENTAN
Grant Number JPMJSN14A1, Japan.




\end{document}